# Adiabatic cross-polarization via intermediate dipolar-ordered state

Jae-Seung Lee and A. K. Khitrin

*Department of Chemistry, Kent State University, Kent OH 44242-0001, USA*

**Abstract**

It is experimentally demonstrated that an adiabatic demagnetization – remagnetization scheme, where the Zeeman order of abundant nuclei is first adiabatically converted into the dipolar order, and then, into the Zeeman order of rare nuclei, can significantly increase polarization of rare nuclei compared to the conventional cross-polarization technique.

Cross-polarization (CP) [1] is a standard technique in solid-state NMR for increasing polarization and, therefore, signals of nuclei with low gyromagnetic ratios. It is based on equalizing polarizations of nuclei of different types. When the number of rare nuclei $N_S$ in a sample is much smaller than the number of abundant nuclei $N_I$, polarization of the rare nuclei can be increased by a factor of $\gamma_I/\gamma_S$, where $\gamma_I$ and $\gamma_S$ are gyromagnetic ratios of the abundant and rare nuclei, respectively. Equalizing polarizations is not the most efficient way to transfer the Zeeman order from abundant to rare spins. It has been known a long time ago that adiabatic transfer of order (polarization) would be more efficient. A thermodynamic consideration, based on conservation of entropy, is straightforward and not presented here. It shows that, compared to the conventional CP scheme, adiabatic transfer of polarization from abundant to rare nuclei can additionally increase polarization of rare nuclei by a factor of $(N_I/N_S)^{1/2}$. Several adiabatic CP schemes have been proposed in [1], and the mechanisms of adiabatic transfer have been further studied theoretically in [2,3]. However, as far as we know, no efficient experimental method has been demonstrated, and the conventional CP [1] remains a standard solid-state NMR technique for more than three decades.

Here we present a simple adiabatic CP scheme (ACP), which creates significantly larger polarization of rare nuclei than the conventional CP. The first pulse of the sequence (Fig. 1), with adiabatic frequency sweep and low RF amplitude, starts irradiation on the abundant nuclei far off-resonance and then, its frequency gradually approaches the center of the spectrum. This pulse converts the Zeeman order of the abundant spins into the dipolar order. Secular dipole-dipole interactions between spins with different gyromagnetic ratios consist of ZZ terms only. However, all the spins have a common reservoir of dipole-dipole interactions [4]. Common spin temperature is established by flip-flops of the abundant spins, which adjust orientations of the abundant spins in the vicinity of rare nuclei according to the existing Z-fields produced by the rare nuclei. This equilibration process does not require flips of the rare spins. The second adiabatic frequency-sweeping pulse is applied to the rare spins. Irradiation starts at the center of the spectrum of rare nuclei and then the offset gradually increases. This pulse converts the dipole-dipole order into the Zeeman order of rare spins.

For the first test of this scheme we have chosen a liquid-crystalline sample with natural isotope abundances, where a high-resolution $^{13}$C spectrum can be acquired without magic-angle spinning of the sample. The experiment has been performed using a Varian Unity/Inova 500 MHz NMR spectrometer. Liquid crystal 5CB has been used as purchased from Aldrich, without further purification. Both $^1$H and $^{13}$C adiabatic frequency-sweeping pulses are shaped pulses with constant RF amplitude and time-dependent phase. The number of constant-phase steps in each of the pulses is 50 K. Each of the two pulses has 100 ms duration and 40 kHz frequency sweeping range. The RF fields' amplitudes ($\gamma B_1/2\pi$) for $^1$H and $^{13}$C pulses are 1.6 kHz and 4.2 kHz, respectively. The spin-lattice relaxation time of the dipole-dipole reservoir $T_{1D}$ was measured to be 0.54 s. The hetero-nuclear decoupling sequence used in all experiments is SPINAL-64 [5].

The results are shown in Fig. 2. The spectrum (a) is acquired with a single $^{13}$C pulse, starting with a thermal equilibrium state. The spectrum (b) is recorded after 2.65 ms CP with the matched Hartmann-Hahn RF fields of 19 kHz amplitudes. The result of the ACP with the pulse sequence in Fig. 1 is shown in Fig. 2(c). The numbers near the peaks in Figs. 2(b) and 2(c) show the relative intensities of those peaks compared to the single-pulse acquisition.

One can see that the $^{13}$C magnetization created by our ACP sequence is significantly higher than the magnetization resulting from the conventional CP. In fact, it is even higher than the theoretical limit of the conventional CP in this system. The experiments on optimizing the ACP scheme for solid samples at magic-angle spinning are now in progress.

The work was supported by the Kent State University.

**Figure captions**

Fig. 1. Pulse sequence.

Fig. 2. $^{13}$C spectra of 5CB at 298 K: (a) single-pulse acquisition; (b) cross-polarization; (c) adiabatic cross-polarization.

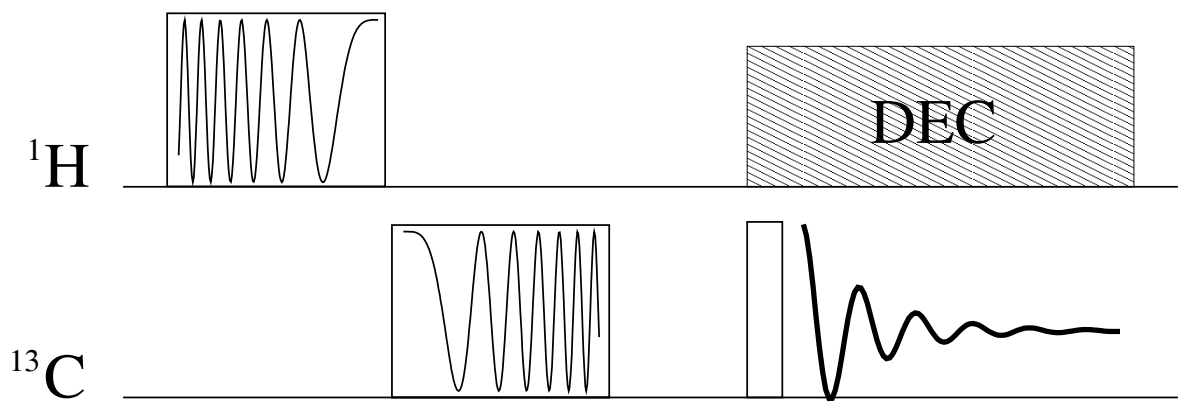

**Fig. 1**

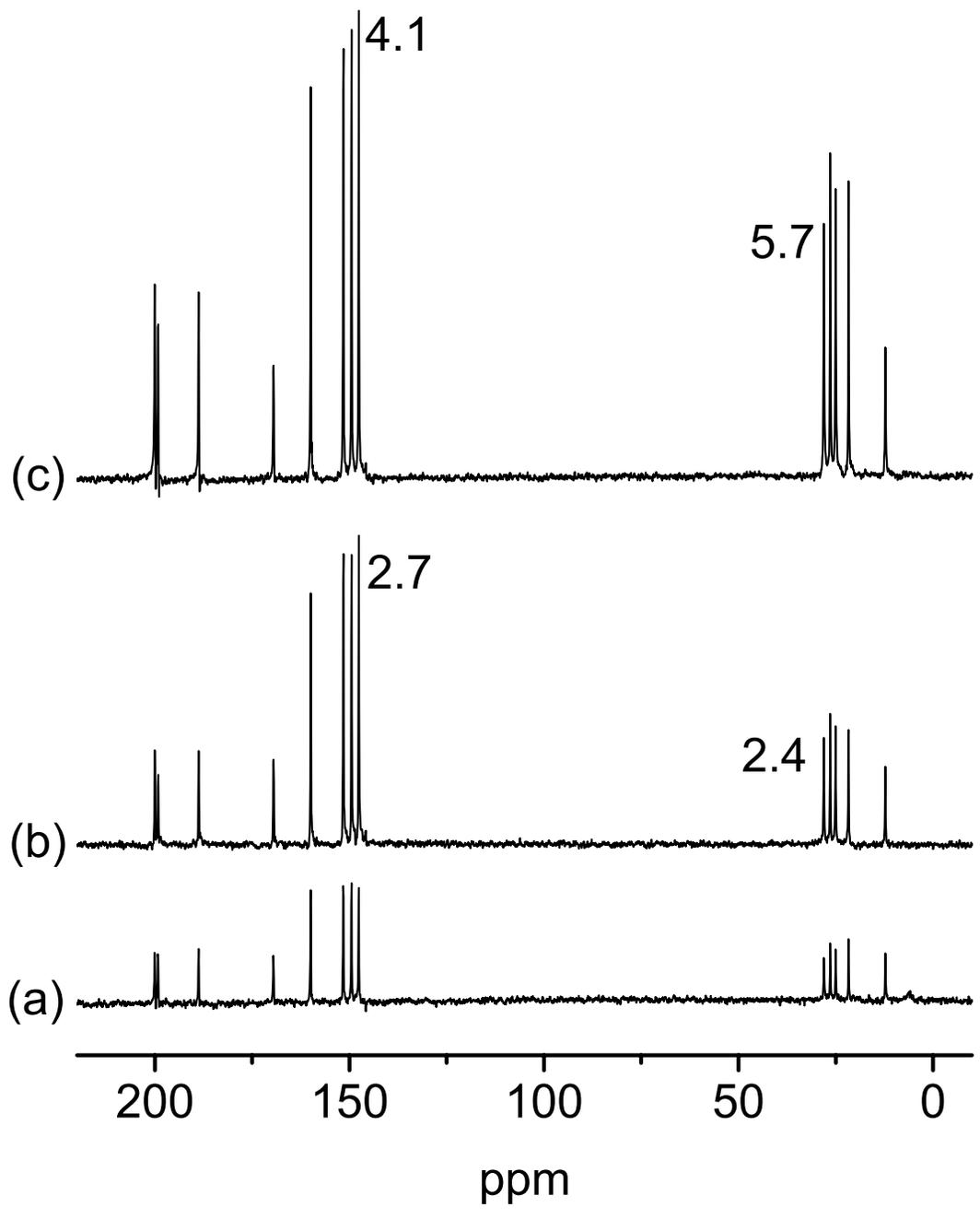

**Fig. 2**